\documentclass[aps,prl,twocolumn,groupedaddress,amsmath,amssymb]{revtex4-1}
\usepackage{graphicx}
\usepackage{color}

\graphicspath{{./}{./Figs/}}

\newcommand{\BraKet}[3]{\left\langle #1 \middle| #2 \middle| #3 \right\rangle}

\newcommand{\be}[1]{\begin{eqnarray}{\label{e#1}}} 
\newcommand{\beq}{\begin{eqnarray}}
\newcommand{\eeq}{\end{eqnarray}} 

\newcommand{\hide}[1]{}
\newcommand{\Eq}[1]{\textcolor{blue}{{Eq.}\!\!~(\ref{#1})}}

\newcommand{\Fig}[1]{\textcolor{blue}{Fig.}\!\!~\ref{#1}}
\definecolor{myred}{rgb}  {0.5,0.0,0.0}

\newcommand{\sect}[1]{{\bf #1.-- }}
\begin{document}

\title{Enforcing Levy relaxation for multi-mode fibers with correlated disorder}

\author{Yaxin Li$^1$}
\author{Doron Cohen$^2$}
\author{Tsampikos Kottos$^1$}

\address{
$^1$\mbox{Wave Transport in Complex Systems Lab, Physics Department, Wesleyan University, Middletown CT-06459, USA}\\
$^2$\mbox{Department of Physics, Ben-Gurion University of the Negev, Beer-Sheva 84105, Israel}
}

%\date{\today}

\begin{abstract}
Environmental perturbations and noise are source of mode mixing and interferences between the propagating modes of 
a complex multi-mode fiber. Typically, they are characterized by their correlation (paraxial) length, and their spectral 
content which describes the degree of coupling between various modes. We show that an appropriate control of these 
quantities allows to engineer Levy-type relaxation processes of an initial mode excitation. Our theory, based on 
Random Matrix Theory modeling, is tested against realistic simulations with multi-mode fibers. 
%The Random Matrix Theory prediction is tested against simulations with multi-mode fibers.
\end{abstract}


\maketitle

%%%%%%%%%%%%%%%%%%%%%%%%%%%%%%%%%%%%%%%%%%%%%%%%%%%%%%%%%%%%%%
\sect{Introduction}
Multi-mode wave dynamics in the presence of noisy environmental perturbations has always been a topic of interest 
for a variety of frameworks ranging from quantum and matter waves to classical waves. In typical circumstances, noise 
is considered an evil, since it degrades the efficiency of the structures employed to perform useful operations on these 
waves. For example, in the frame of quantum electronics, optics and matter wave physics noise is responsible for 
decoherence effects, being damaging to emerging quantum information and computation technologies \cite{S08,DM03,
AO14,ARS01}. In a similar manner, in classical wave technologies (e.g optics, microwaves or acoustics), noise pollutes 
the signal carried by a propagating wave, thus degrading the transfer of information \cite{B11,T02,NK03}. It is perhaps 
for this particular reason that researchers have explored, over the years, various strategies to eliminate noise sources. 
An alternative approach would be to utilize its presence, and via appropriate design use it in order to control the signal propagation.

An example were such strategy might be useful, comes from the area of fiber communication and data processing 
where single-mode fibers have reached their limitation as far as information capacity is concerned \cite{K00,KM17}. 
Instead, multi-mode fibers (MMF) and/or multi-core fibers (MCF) offer new exciting opportunities since their modes 
can be used as extra degrees of freedom for carrying additional information - thus increasing further the information 
capacity \cite{KM17,HK13}. Unfortunately, MMF suffer from mode coupling due to environmental perturbations (index 
fluctuations, fiber bending etc), occurring along the paraxial direction of propagation which both cause crosstalk and 
interference between propagating signals in different modes \cite{HK13,RASC14,XABRRC16,LCK19,r2,r3,r4,r5}. 
Under such conditions, one typically expects fast relaxation of an initial mode excitation in the mode-space and thus 
degradation of the information carried by the signal.

%%%%%%%%%%%%%%%%%%%%%%%%%%%%%%%%%%%%%%%%%%%%%%%%%%%%%%%%%%%%%%
\sect{Outline} 
Below we develop a protocol that controls the relaxation process of an initial mode excitation 
towards its ergodic limit $\sim 1/N$, where $N$ is the dimensionality of the mode space. The mode mixing 
is due to quenched disorder associated with external perturbations along the propagation direction $z$ of a MMF. The 
proposed scheme is based on the manipulation of these dynamical perturbations that are responsible for the mixing 
between the various modes. They are characterized by a spectral exponent~${s\in[-1,1]}$ and by a correlation length~$z_c$. 
We show that the relaxation process is given by a generalized Levy-law with a power exponent ${\alpha \in [0,2]}$  
that is dictated by the spectral content of the deformation. 
The generic prediction is based on Random Matrix Theory (RMT) modelling, 
and then tested against simulations with a MMF. 
In the latter context we provide a simple recipe for engineering the spectral content of the external perturbations, 
by designing the roughness of the fiber cross-section. Although our analysis refers  
to MMFs, the proposed methodology will have ramifications in other fields, ranging from decoherence 
management in quantum dots, matter waves and quantum biology \cite{ECRAMCBF07,PHFCHWBE10,SIFW10,L11,E11,
FSC11}, to acoustics and control of mechanical vibrations \cite{B11,T02,NK03}.

%%%%%%%%%%%%%%%%%%%%%%%%%%%%%%%%%%%%%%%%%%%%%%%%%%%%%%%%%%%%%%
\sect{Modeling} 
We assume that imperfections induce coupling only between forward propagating modes (paraxial approximation). 
Furthermore, these perturbations vary with the propagation distance~$z$. 
The $z$-dependent Hamiltonian $H$ that describes the field propagation along the MMF 
can be written as ${H=H_0+V(z)}$ where $H_0$ describes the unperturbed fiber, 
and $V(z)$ is a spatial-dependent potential characterized by a correlation 
function $\langle V(z')V(z'')\rangle={\cal C}[(z'-z'')/z_c]$. 
We assume that the correlation function does not have heavy tails, 
and therefore, in practice, the fiber can be regarded as a chain of~$L$  
uncorrelated segments, indexed by ${t=1\cdots L}$.     
The paraxial Hamiltonian that describes the field propagation within the $t$ segment is 
$H^{(t)}=H_0+ \varepsilon B^{(t)}$, 
where $\varepsilon$ indicates the strength of the perturbation. 
In the mode-basis of the unperturbed system, 
the $N\times N$ matrix $H_0$ is diagonal, 
with elements $\left(H_0\right)_{nm}=\delta_{nm} \beta_n$, 
where $\beta_1<\beta_2<\cdots\beta_N$ are the propagation constants 
of the modes $n=1,2,\cdots,N$. 
In this basis, the constant perturbation matrix $B^{(t)}=\left(B^{(t)}\right)^{\dagger}$ is responsible for the mode mixing. 
We assume that the perturbations associated with each concatenated segment are uncorrelated with one-another 
i.e. $\langle B^{(t)}B^{(t')}\rangle = \delta_{t,t'}$. 

The field propagation in each section is described by the unitary matrix 
\begin{equation}
% \ket{\Psi^{(t)}} = U^{(t)}\ket{\Psi^{(t-1)}},\quad 
U^{(t)}=e^{-i\left(H_0+ \varepsilon B^{(t)}\right) z_c} 
\label{Tmatrixk}
\end{equation}
%
% where $\ket{\Psi^{(t)}}$ is the propagated field. 
The modal field amplitudes ${\Psi_n(z)\equiv\langle n|\Psi^{(t)}\rangle}$ 
at distance ${z= t\times z_c}$ 
along the MMF are obtained by operating on the initial state 
${\Psi_n(0)=\delta_{n,n_0}}$ with a sequence of $U^{(t)}$ matrices.  
This multi-step dynamics generates a distribution ${P_z(n|n_0)=|\Psi_n(z)|^2}$. 

\begin{figure}
\includegraphics[width=0.5\textwidth]{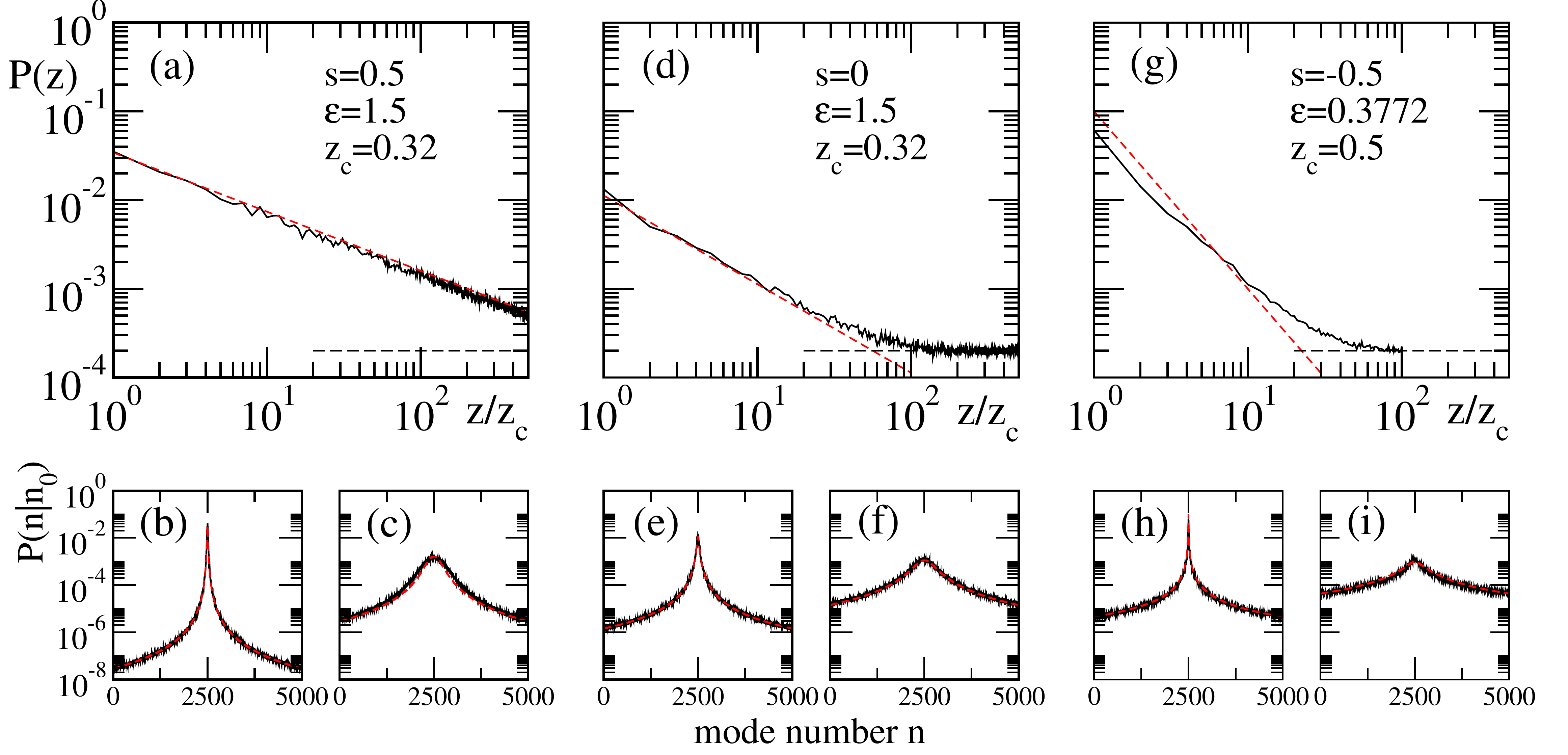}
\caption{
{\bf Power law relaxation (RMT).} 
We consider here an RMT-modelled propagation of a pulse 
in a multi-mode fiber with ${N =5000}$ modes. 
The results are averaged over more than 50 disorder realizations. 
The upper row shows the decay of the survival probability for 
bandprofile with (a)~$s=0.5$; (d)~$s=0$; and (g)~$s=-0.5$. 
The values of $\epsilon$ and $z_c$ are indicated in the figures.
The decay is described by a power-law (red line) 
with power exponents ${\alpha=s{+}1}$,  given by \Eq{genpower}. 
The black dashed lines indicate the ergodic limit $1/N$. 
Panels~(b,e,h): the waveprofile after a propagation distance $z=z_c$. 
The red dashed lines represent the theoretical prediction \Eq{W1}. 
Panels~(c,f,i): the waveprofile for ${z/z_c = 100, 10, 10}$ respectively.
In all cases, the black solid lines represent simulation data 
and the red dashed lines represent the theoretical predictions.}
\label{fig1}
\end{figure}

We characterize the spectral content of the perturbations 
via the lineshape of the band profile. 
Specifically we consider below an {\em engineered} bandprofile 
that is characterized by power-law tails     
\begin{equation}
\label{bpf}
|B_{n,m}|^2 \ \sim \ \frac{2}{|n-m|^{s}}, 
\ \ \ {\rm where}\,\,\, s\in[-1,1]. 
\end{equation}
RMT considerations (that we outline below) 
imply that for large correlation length, 
the relaxation process is formally like a Levy-flight, 
for which the survival probability exhibits 
a power-law decay  
\begin{equation}
\label{sp}
\mathcal{P}(z) \ \equiv \ \overline{P_z(n_0|n_0)} \ \sim \ \left(\frac{1}{z}\right)^{1/\alpha};\,\, {\rm where}\,\, \alpha=1+s. 
\end{equation}
The overbar indicates an average over different initial conditions or/and disorder realizations. 
An RMT demonstration of this decay is provided in \Fig{fig1}, 
and further discussed later.

%--------------------------------------------------
% RMT long zc
%--------------------------------------------------

%%%%%%%%%%%%%%%%%%%%%%%%%%%%%%%%%%%%%%%%%%%%%%%%%%%%%%%%%%%
%%%%%%%%%%%%%%%%%%%%%%%%%%%%%%%%%%%%%%%%%%%%%%%%%%%%%%%%%%%
\sect{Engineered Bandprofile} 
We consider a cylindrical fiber with core radius $a$. 
We assume TM propagating waves in the fiber. 
The solutions of the Helmholtz equation, 
under the requirement that the eigenmodes are finite at $r\rightarrow 0$, take the form
\beq
\label{unpert1}
\Psi^{\nu,\ell}(r,\theta,z)=C_{\nu,\ell} \ J_{\nu} \left(  \frac{x_{\nu,\ell}}{a} r  \right) \  e^{i\nu\theta} \ e^{i \beta_{\nu,\ell} z}
\eeq
where $\Psi(r,\theta,z)$ is the electric field of the TM mode, $\nu$ is the azimuthal mode index, $\ell$ is the radial 
mode index and $J_{\nu}(\cdot)$ is the first-type Bessel function of order $\nu$. The argument $x_{\nu,\ell}$ indicates 
the zeroes of the Bessel functions and for simplicity we only considered right-hand polarization. In case of a metal
-coated core, the electric field satisfies the boundary conditions $\Psi(a,\theta,z) = 0$ and $\frac{\partial}{\partial
\theta}\Psi(a,\theta,z)= 0$. These lead to the following quantization for the propagation constants for forward 
propagating waves in the $(\nu,\ell)$ channel 
\beq
\label{unpert2}
\beta_{\nu,\ell} \ \ = \ \ \sqrt{ \left( \frac{\omega}{c} \right)^2 -  \left( \frac{x_{\nu,\ell}}{a} \right)^2}
\eeq  
where $c$ is the speed of light in the fiber medium. In our simulations below, 
the core-index is ${\bar n}=1.5$,  
and the incident light has $\lambda_{vac} = 1.55 \mu$m. 

\begin{figure}
\includegraphics[width=0.4\textwidth]{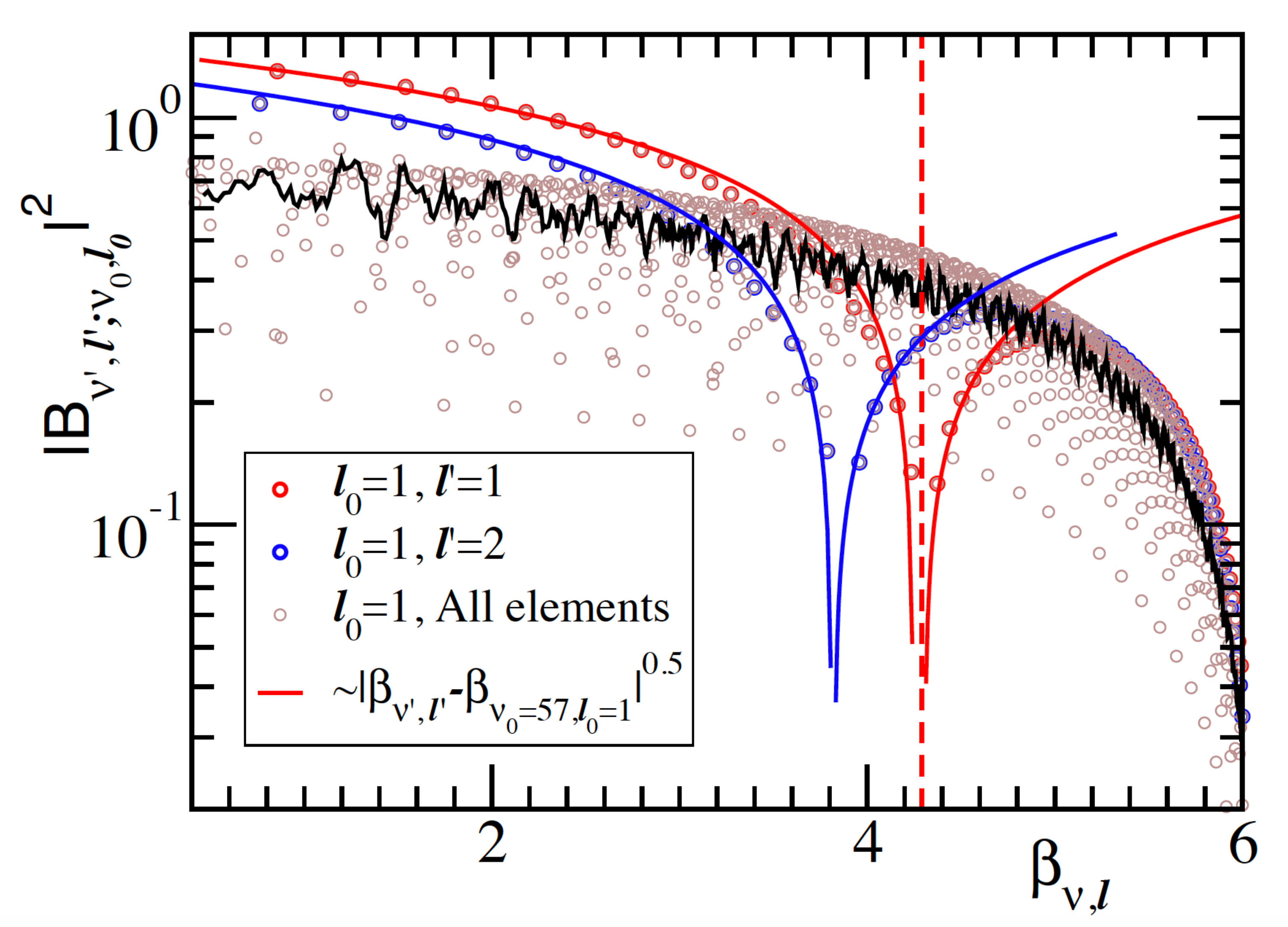}
\caption{
{\bf The band profile.} 
The matrix elements $|B_{\nu\ell,\nu_0\ell_0}|^2$ 
for an $N{=}1031$ MMF with core radius $a=15\mu$m, 
calculated using \Eq{Bmat} for $s=-0.5$. 
The red symbols are for the ${\ell=\ell_0=1}$ block.
The other symbols are the couplings to the ${\ell \ne \ell_0}$ modes.
The solid lines indicate the $\propto |\nu-\nu_0|^{-s}$ lineshape  
that is implied by \Eq{bnm}.  
If all the $\ell$~modes are included (brown symbols)
the averaged bandprofile (black line) 
features ${s_{\text{eff}}=0}$ lineshape.}
\label{fig2}
\end{figure}

Next, we enforce a pre-designed mode-mixing between the modes of the cylindrical (perfect) fiber via engineered 
boundary deformations. It turns out that a boundary deformation $D(\theta)$ due to e.g. surface roughness of the 
fiber, is related to the perturbation matrix $B_{nn'}$ via a ``wall'' formula \cite{C00,BCH00}
\beq
\label{wall}
B_{n,n'} = \oint  \partial\psi_{n}(\theta) \ \partial\psi_{n'}^*(\theta)  \ 
D(\theta) d\theta  
\eeq
where the sub-index indicates $n\equiv(\nu,\ell)$. In \Eq{wall} the integration is along the boundary, and 
$\partial$ indicates the normal derivative. Substitution of the eigenmodes of the unperturbed fiber \Eq{unpert1}
in \Eq{wall} leads to the following expression for the matrix elements of the $B$-matrix     
\beq \label{Bmat}
\BraKet{\nu,\ell}{B}{\nu',\ell'} \ \ = \ \ c_{\nu,\ell} c_{\nu',\ell'} \ b_{\nu,\nu'}
\eeq
where $c_{\nu,\ell} = \ C_{\nu,\ell} \ J_{n}'(\frac{x_{\nu,\ell}}{a}r)|_{r=a}$. The matrix elements within each 
$\ell$-block are essentially the FT of the azimuthal deformation $D(\theta)$ i.e.
\beq
\label{block}
b_{\nu,\nu'} \ \ = \ \  \oint  D(\theta) e^{i(\nu-\nu')\theta} \frac{d\theta}{\pi}
\eeq
From \Eq{block} it is straightforward to identify the appropriate deformations $D(\theta)$ whose FT lead 
to perturbation matrices $B$ with a desired power-law bandprofile, see \Eq{bpf}. A simple recipe is to 
introduce $D(\theta)$ as a cosinus-Fourier series with coefficients that are random phases \cite{FS97}. Namely, 
\beq
D(\theta) \ \ = \ \ \sum_{q=1}^N \sqrt{Q_q} \ \cos(q \theta+\varphi_q)
\eeq
with ${ Q_q =q^{-s} }$, 
while $\varphi_q$ is a random phase in the interval $[0,2\pi)$. 
Substitution of the above deformation families 
back to \Eq{block} leads to the following band-profile for the $\ell$-th block sub-matrix
\beq \label{bnm}
|b_{\nu,\nu'}|^2  \ \ = |Q_{|\nu-\nu'|}| \ \ = \ \ \frac{1}{|\nu-\nu'|^s}
\eeq
The band profile that is implied by \Eq{Bmat} with \Eq{bnm} is illustrated in \Fig{fig2} 
for an MMF with core radius $a=15\mu$m. We have considered perturbations with $s=-0.5$. 
If we keep a single $\ell$ block (say $\ell_0{=}1$) the band profile features power-law tails as in \Eq{bpf}.
But if we include all the $\ell$ modes, we get an averaged bandprofile that is not singular 
at ${\beta_{\nu,\ell} \sim \beta_{\nu_0,\ell_0}}$, meaning an effective exponent ${s_{\text{eff}}=0}$.

%--------------------------------------------------
% RMT long zc
%--------------------------------------------------

%%%%%%%%%%%%%%%%%%%%%%%%%%%%%%%%%%%%%%%%%%%%%%%%%%%%%%%%%%% 
\sect{RMT framework} 
The disordered nature of the perturbations allows RMT modeling for the matrix $B^{(t)}$, see \cite{HK13,LCK19,r3,r4}. 
Within the RMT framework, the matrix elements are not calculated but generated artificially. 
Specifically, we assume that the elements are uncorrelated random numbers drawn 
from a normal distribution centered at zero. We further assume, {\em dogmatically}, that they have variance 
in accordance with the power law lineshape of \Eq{bpf}, and that the mode propagation constants 
are equally spaced, namely, ${\beta_n=n \Delta}$, where ${n=1,\cdots,N}$. 
We will use the RMT modeling as a benchmark against which we shall compare 
the results from MMF simulations that use {\em engineered} bandprofiles.       

The correlation length $z_c$ plays a major role in the decay of the survival probability.
It has been argued \cite{LCK19} on the basis of analogy with Fermi-golden-rule that the strength~$\varepsilon$  
of the perturbation translates into a length scale ${ z_{\Gamma}= ( \varepsilon^2/\Delta )^{-1} }$.    
If we have ${z_c < z_{\Gamma}}$ there is an intermediate stage during which the decay 
is exponential ${ \mathcal{P}(z) \propto e^{-\lambda z} }$, see details in \cite{LCK19}.  
Otherwise the decay is power law (see details below). 
Stand alone power-law decay for ${z > z_c > z_{\Gamma}}$ is demonstrated in \Fig{fig1}
for representative values of $s$.   

%------------------------------------------------------
% MMF long zc
\begin{figure}
\includegraphics[width=0.5\textwidth]{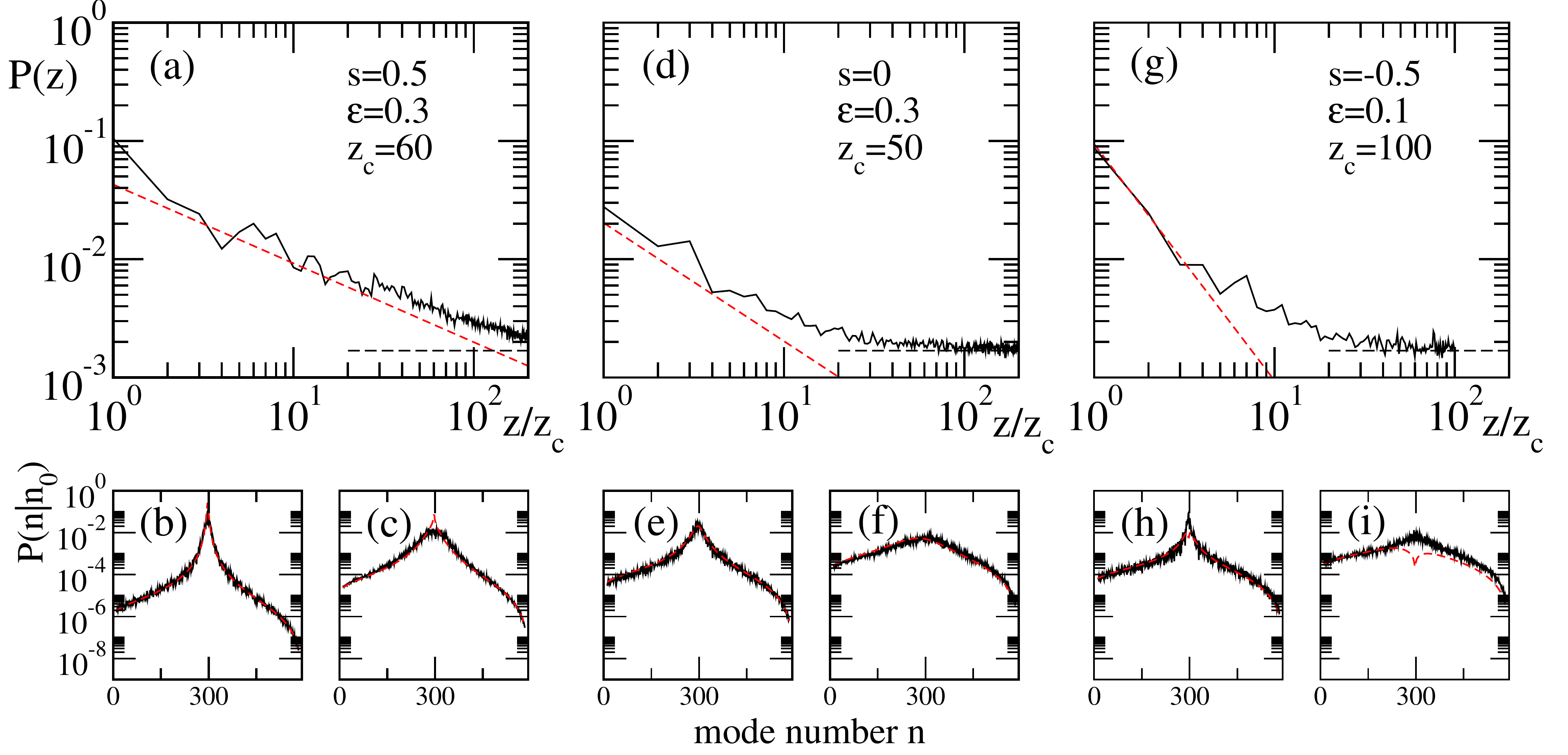}
\caption{
{\bf Relaxation for a graded-index ring-core MMF.} 
The MMF has core radius $a=100 \mu$m, and $N=593$ modes. 
All the modes are confined within the ${\ell=1}$ subgroup. 
The results are averaged over initial conditions and realizations of the rough surface. 
The decay of the survival probability follows the power-law prediction \Eq{genpower} 
with (a)~$s=0.5$; (d)~$s=0$; and (g)~$s=-0.5$ (red dashed lines). 
Panels~(b,e,h): the waveprofile after a propagation distance ${z=z_c}$.
Panels~(c,f,i): the waveprofile for ${z/z_c = 10, 5, 5}$ respectively.
The solid lines represent simulation data and the red  dashed lines represent theoretical predictions.
} 
\label{fig3}
\end{figure}
%------------------------------------------------------
%

Following an argument that extends first-order perturbation theory \cite{LCK19,CIK00,CK01,HKG09} 
the spreading kernel will saturate, for long enough $z_c$,  
to a generalized-Lorentzian line shape $\text{Prob}(n|n_0)=W(n-n_0)$ that is characterized 
by a power law tail, namely,  
\beq \label{gL}
W(n{-}n_0) \ \sim \ \frac{\varepsilon^2|B_{n,n_0}|^2}{[(n-n_0)\Delta]^2} 
\ = \ \frac{(\varepsilon/\Delta)^2}{|n-n_0|^{1+\alpha}}
\eeq
where ${\alpha=1+s}$. In order to calculate the spreading profile after a propagation distance~$z$, 
one has to perform $t=z/z_c$ successive convolutions. If the kernel has finite second-moment, 
which is the case for ${\alpha>2}$, one obtains Gaussian spreading, as implied by the central limit theorem. 
This would lead to ${P(z) \propto \sqrt{1/z} }$ decay of the survival probability. 
Our interest below is in ${0<\alpha<2}$, which leads to an anomalous so-called Levy-process.

For the purpose of successive convolutions, the $W(r)$ kernel can be approximated 
as the FT of $w(\kappa) = \exp(-|\gamma \kappa|^{\alpha})$, 
where $\gamma$ is a fitting constant. 
Generally, these type of distributions do not have analytic expressions.
However, assuming that the kernel obeys one-parameter scaling,  
it follows from normalization that 
\beq
\gamma \ \ = \ \ C \left(\frac{\varepsilon}{\Delta}\right)^{2/\alpha}
\eeq
where $C$ is a numerical prefactor of order unity. 
Successive convolutions lead to the Levy $\alpha$-stable distribution.
Namely, after $t$ steps $\text{Prob}_t(n|n_0)$ will be the FT    
of $[w(\kappa)]^t$. This means that the width parameter evolves 
as ${\gamma \mapsto \gamma t^{1/\alpha}}$. 
The survival probability is simply the area of $\exp(-t|\gamma \kappa|^{\alpha})$, 
and accordingly we get
\beq
\mathcal{P}(z) \ \ = \ \ \frac{1}{\gamma} \left(\frac{z_c}{z}\right)^{1/\alpha} 
\eeq
Note the special cases ${\alpha=1,2}$ that correspond 
to Lorentzian spreading and Normal diffusion respectively.

%------------------------------------------------------
% MMF short zc
\begin{figure}
\includegraphics[width=0.5\textwidth]{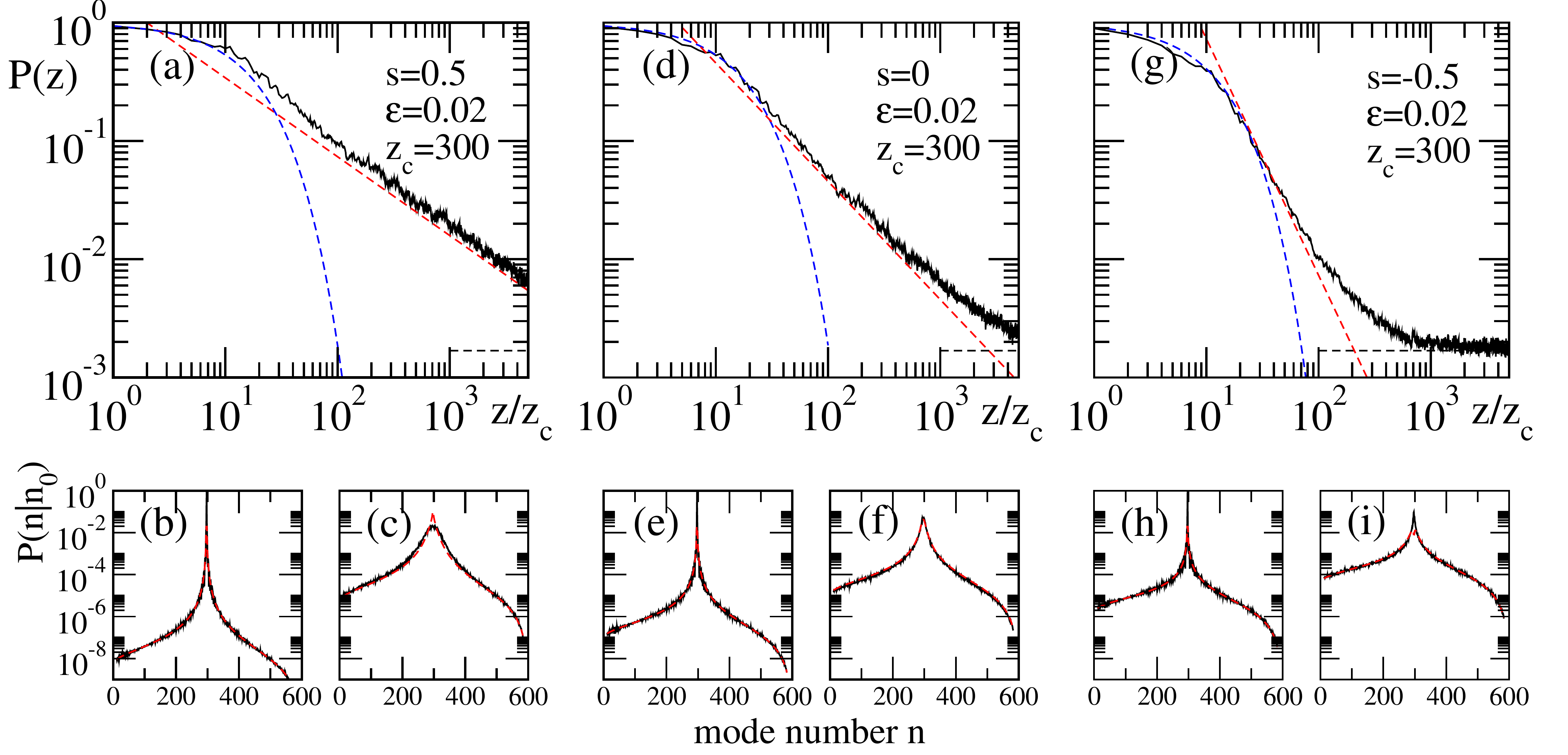}
\caption{
{\bf An exponential relaxation stage.} 
The same as in \Fig{fig3} for disorder that features ${ z_c < z_{\Gamma} }$.    
Consequently an intermediate exponential stage appears (blue lines).  
Panels~(b,e,h): the waveprofile after a propagation distance ${z=z_c}$.
Panels~(c,f,i): the waveprofile for ${z/z_c = 1000, 100, 25}$.
} 
\label{fig4}
\end{figure}
%------------------------------------------------------

%%%%%%%%%%%%%%%%%%%%%%%%%%%%%%%%%%%%%%%%%%%%%%%%%%%%%%%%%%% 
\sect{Generalized Lorentzian} 
In practice we have to handle in our formulas the whole bandprofile.  
In the standard case $\alpha{=}1$, and it is common to fit \Eq{gL} into a Lorentzian.
In general  $\alpha{\ne}1$, and we use for fitting a Generalized Lorentzian~(GL), namely, 
\beq \label{W1}
W(n{-}n_0) \ &\approx& \ \frac{1}{\pi} \frac{\gamma^{\alpha}}{|n-n_0|^{1{+}\alpha}+\gamma^{1{+}\alpha}}
\\ \label{Kernelgen}
&\mapsto& \ \ \frac{2\varepsilon^2|B_{n,n_0}|^2}{|\beta_n-\beta_{n_0}|^2+\Delta^{1-\alpha}\Gamma_{\rm GL}^{1+\alpha}}
\eeq
where in the second line we have manipulated the second term in the denominator:  
this term acts as a regularization parameter for the ${n\sim n_0}$ singularity, 
while at the tails it is negligible; 
therefore, for numerical purpose, only its near-diagonal value is important.   
For a smooth bandprofile the regularization parameter is   
\beq
\label{Gamma_GL}
\Gamma_{\rm GL} \ = \  \gamma \Delta \ = \ \left(\frac{2 \pi \varepsilon^2}{\Delta^{2{-}\alpha}}\right)^{1/\alpha}
\eeq
and 
\beq \label{genpower}
\mathcal{P}(z) \ \ = \ \ \frac{1}{\pi}\frac{\Delta}{\Gamma_{\rm GL}} \left(\frac{z_c}{z}\right)^{1/\alpha};\quad \alpha=1+s 
\eeq
The convention we use for $\Gamma_{\rm GL}$ is such that for ${\alpha=1}$ 
the Lorentzian (ballistic-like) decay of \cite{LCK19} is recovered. 
For a general bandprofile we determine $\Gamma_{\rm GL}$ via normalization of \Eq{Kernelgen}. 
We demonstrate in all our figures that this practical numerical procedure is a valid approximation.

%------------------------------------------------------
% MMF ungraded-index 
\begin{figure}
\includegraphics[width=0.4\textwidth]{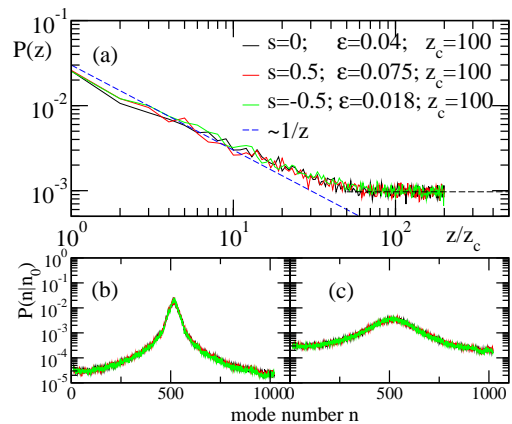}
\caption{
{\bf Relaxation for an ungraded core-index MMF.} 
The $N=1031$ modes for different $\ell$-groups are mixed together via the perturbation matrix $B$. 
A ballistic decay with $\alpha=1$ (blue dashed line) is observed irrespective of the $s$-value.
The core radious is $a=15 \mu$m. } 
\label{fig5}
\end{figure}
%------------------------------------------------------

%%%%%%%%%%%%%%%%%%%%%%%%%%%%%%%%%%%%%%%%%%%%%%%%%%%%%%%%%%%%%%
\sect{Levy relaxation for a graded-index MMF} 
In order to design a MMF that features the desired power law lineshape of \Eq{bpf}, 
we have to effectively eliminate transitions to ${\ell \ne \ell_0}$ modes, where $\ell_0$ indicates 
the initial radial excitation. The latter constraint is achieved in case of graded-index MMFs \cite{FGGJPJKABSRWW16}. 
The grading of the refraction index is like a radial potential. 
Such potential does not affect the form of \Eq{Bmat}, 
but it does shift ``horizontally" the branches of \Eq{unpert2} 
that are illustrated in \Fig{fig2}.     
Thus, effectively, only modes that belong 
to the same radial mode-group (${\ell=\ell_0}$) are mixed,
with perturbation matrix that is proportional to $b_{\nu,\nu'}$ of \Eq{bnm}.
  
In our computational example we chose to focus on a low-$\ell$ radial group (e.g. $\ell=1$) in 
order to minimize the effects of radiative losses. 
In \Fig{fig3} we have considered three representative values of~$s$. 
Power law decay is demonstrated for a disordered MMF that has a correlation scale $z_c>z_{\Gamma}$. 
The survival probability follows a Levy-type relaxation given by \Eq{genpower} 
with powers ${\alpha=1+s}$. Moreover, the probability distribution 
can be approximated by the generalized Lorentzian of \Eq{Kernelgen}.   

For completeness we have performed simulations for a MMF with disorder that has a correlation length ${z_c < z_{\Gamma}}$, 
see \Fig{fig4}. As expected form the analysis in \cite{LCK19}, an intermediate exponential stage appears. 
This exponential decay reflects Fermi's golden rule. However, the long time decay becomes power law, 
as in \Fig{fig3}, until the waveform reaches an ergodic distribution.

%%%%%%%%%%%%%%%%%%%%%%%%%%%%%%%%%%%%%%%%%%%%%%%%%%%%%%%%%%%%%%
\sect{Universal relaxation} 
We turn to consider an ungraded-index MMF. 
Transitions between modes that correspond to different $\ell$-groups cannot be neglected, 
and therefore the bandprofile of the perturbation matrix $B$ becomes effectively flat, see \Fig{fig2}, 
irrespective of the spectral content of the deformation $D(\theta)$.
Namely, the information about~$s$ is washed out once the propagation constants 
are ordered by magnitude and the $\nu$ indices are shuffled. 
Consequently, the survival probability ${\cal P}(z)$ for large correlation lengths 
(and large propagation distances) decays with ${\alpha_{\text{eff}}=1}$ irrespective of the value of~$s$.  
The numerical demonstration is displayed in \Fig{fig5}.

%%%%%%%%%%%%%%%%%%%%%%%%%%%%%%%%%%%%%%%%%%%%%%%%%%%%%%%%%%%%%%
\sect{Summary} 
We have demonstrated exponential and power law decay for pulse propagation in MMF. 
General $\alpha$ Levy decay can be engineered for graded-index ring-core MMF, 
while universal $\alpha_{\text{eff}}{=}1$ is observed for MMFs with ungraded-index.

%%%%%%%%%%%%%%%%%%%%%%%%%%%%%%%%%%%%%%%%%%%%%%%%%%%%%%%%%%%%%%
%%%%%%%%%%%%%%%%%%%%%%%%%%%%%%%%%%%%%%%%%%%%%%%%%%%%%%%%%%%%%%
%%%%%%%%%%%%%%%%%%%%%%%%%%%%%%%%%%%%%%%%%%%%%%%%%%%%%%%%%%%%%%
\clearpage

\sect{Acknowledgments} 
(Y.L) and (T.K) acknowledge partial support by grant No 733698 from Simons Collaborations in MPS, and by an NSF grant 
EFMA-1641109. (D.C.) acknowledges support by the Israel Science Foundation (Grant No. 283/18).

\clearpage

\end{document}